\documentclass[prl,twocolumn,floatfix,showpacs,superscriptaddress]{revtex4}
\usepackage{amsmath}
\usepackage{graphicx}

\begin{document}
 
\title{Finite doping signatures of the Mott transition in the two-dimensional Hubbard model}
\author{G. Sordi} 
\affiliation{D\'epartement de physique and Regroupement qu\'eb\'equois sur les mat\'eriaux de pointe, Universit\'e de Sherbrooke, Sherbrooke, Qu\'ebec, Canada J1K 2R1}
\author{K. Haule}
\affiliation{Department of Physics \& Astronomy, Rutgers University, Piscataway, NJ 08854-8019, USA}
\author{A.-M. S. Tremblay}
\affiliation{D\'epartement de physique and Regroupement qu\'eb\'equois sur les mat\'eriaux de pointe, Universit\'e de Sherbrooke, Sherbrooke, Qu\'ebec, Canada J1K 2R1}
\affiliation{Canadian Institute for Advanced Research, Toronto, Ontario, Canada, M5G 1Z8}

\date{\today}

\begin{abstract}

Experiments on layered materials call for a study of the influence of short-range spin correlations on the Mott transition. To this end, we solve the cluster dynamical mean-field equations for the Hubbard model on a plaquette with continuous-time quantum Monte Carlo. The normal state phase diagram as a function of temperature $T$, interaction strength $U$ and filling $n$ reveals that upon increasing $n$ towards the insulator, there is a surface of first-order transition between two metals at non-zero doping. For $T$ above the critical end line there is a maximum in scattering rate.

\end{abstract}
\maketitle


Simple models in physics can have surprisingly rich sets of solutions that explain complex behavior observed in nature with minimal assumptions. 
The Hubbard model is one such model.
It contains a kinetic energy term that represents the band structure, and an on-site repulsion.
That model has helped understand for example the Mott transition, that is the metal insulator transition induced by strong screened Coulomb interactions between electrons, a fundamental phenomenon in condensed matter physics \cite{mott,ift}.
Several transition metal oxides, for example, are close to a Mott transition and their properties can be explained by the Hubbard model using single-site dynamical mean-field theory (DMFT) \cite{rmp,kotliarRMP,limelette}.

However, for layered doped Mott insulators, such as high temperature superconductors, or for layered organics that show unusual Mott critical behavior \cite{Kagawa:2005}, it is known that single-site DMFT fails because it does not take into account short-range spin correlations. 
Cluster generalizations of DMFT \cite{maier, kotliarRMP} are computationally expensive but they cure this shortcoming.
Recent studies have explored the interaction-driven Mott transition using cluster methods, indeed revealing sharp modifications to the single-site DMFT picture \cite{phk, balzer}. 
However, an overall perspective on the transition induced by doping is missing since previous work has focused on large interaction strength \cite{kyung, hauleDOPING, jarrellNFL, werner8, michelEPL, liebsch}. 

\begin{figure}[!ht]
\centering{
  \includegraphics[width=0.9\linewidth,clip=]{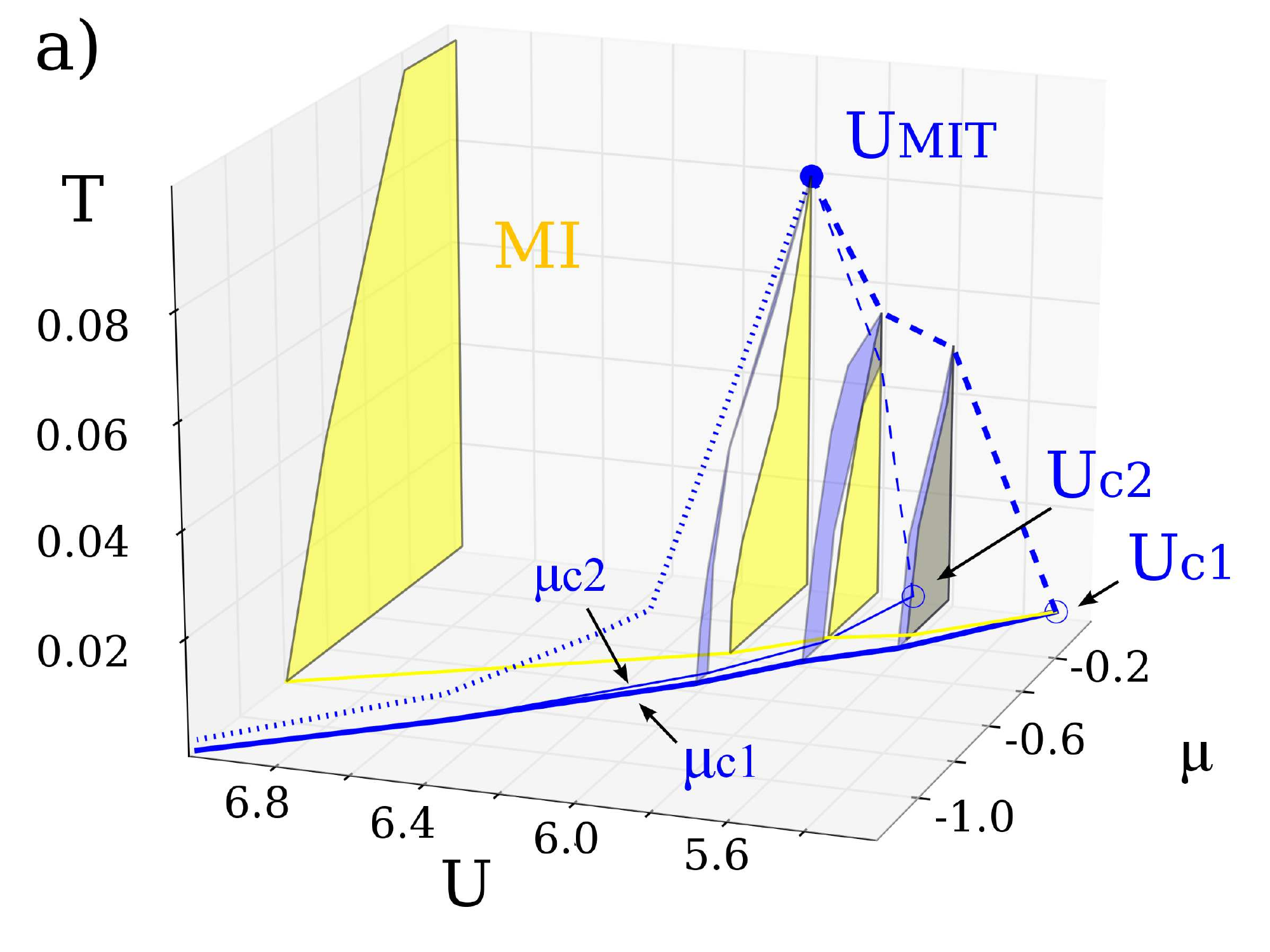}
  \includegraphics[width=0.9\linewidth,clip=]{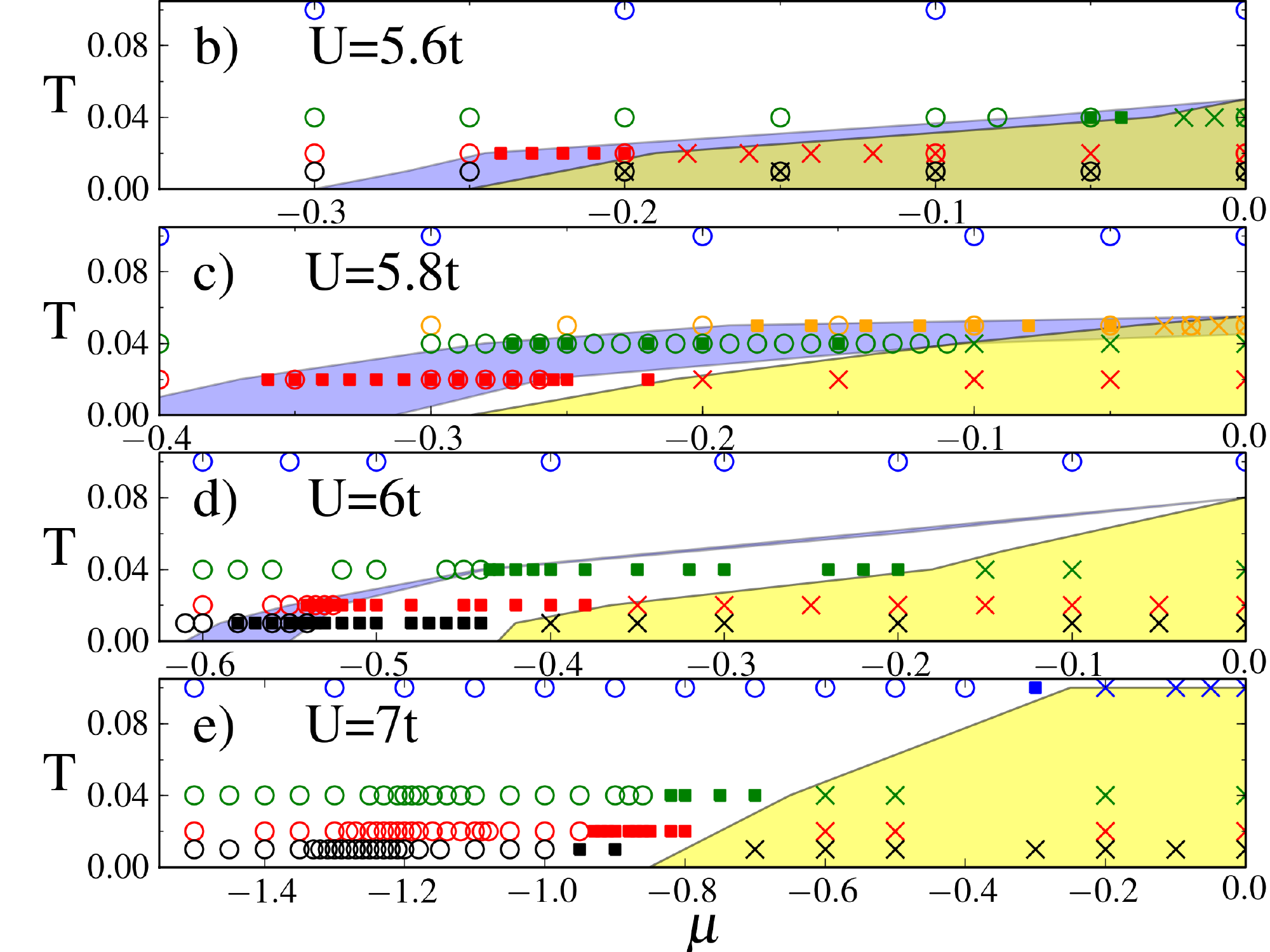}
  }
\caption{(a): Chemical potential-interaction-temperature phase diagram of the two-dimensional Hubbard model obtained by cluster DMFT. Cross-sections at constant $U$ are shown. Dark-gray (blue) shaded regions represent the coexistence of two phases. Light-gray (yellow) areas denote the onset of the Mott insulator state, characterized by a total occupation $n=1$. Projections on $\mu=0$, and $T=0$ planes are also shown (full lines and dashed lines respectively). Open dots mark the extrapolated $T=0$ values of $U_{c1}$ and $U_{c2}$. 
A critical end line (dotted line) begins at the full dot $U_{MIT}$. 
(b),(c),(d),(e): the cross-sections at constant $U$ are shown in several $T-\mu$ plots. Three phases can be distinguished: metal (open circles), Mott insulator (crosses), momentum selective phase where some regions of the Brillouin zone are gapped and others gapless (filled squares). 
}
\label{fig1}
\end{figure}

Here we report the whole normal-state phase diagram of the two-dimensional Hubbard model as a function of chemical potential $\mu$, temperature $T$ and interaction strength $U$. Our results are summarized in Fig.\ref{fig1} that we shall explain below. 
This phase diagram could also be explored with optical lattices of cold atoms where $U$, $T$ and $n$ can be controlled \cite{Jordens:2008, Eckardt:2009}. 
In short, for sufficiently large $U$, there is a first-order phase transition between two paramagnetic metallic phases, one of which evolves continuously from the Mott insulator. There is a large scattering rate associated with that transition. This maximum in scattering extends above the critical end line that terminates the first order surface at progressively lower $T$ and larger doping with increasing $U$. The broad maximum in the scattering rate is centered around what is optimal doping for the high-temperature superconductors. 

{\it Model and method:}
The Hamiltonian for the two-dimensional Hubbard model reads
\begin{equation}
  H = -\sum_{ij\sigma} t_{ij} c_{i\sigma}^\dagger c_{j\sigma}
  + U \sum_{i}  \left(n_{i \uparrow }-\frac{1}{2}\right) \left(n_{i \downarrow }-\frac{1}{2}\right) 
  - \mu\sum_{i} n_{i},
\label{eq:HM}
\end{equation}
where $c_{i\sigma}$ and $c^+_{i\sigma}$ operators annihilate and create electrons on site $i$ with spin $\sigma$, and $n=c^+_{i\sigma}c_{i\sigma}$. $t$ is the hopping amplitude between nearest neighbors, $U$ is the energy cost of double occupation at each site and $\mu$ is the chemical potential. 

We solve this model using cluster dynamical mean-field theory. 
The lattice problem (\ref{eq:HM}) is mapped onto a $2\times2$ plaquette immersed in a bath that is determined self-consistently in such a way that infinite lattice and plaquette have the same self-energy. 
The partition function of the plaquette coupled to the bath is given by \cite{kotliarRMP} 
\begin{equation}
  Z = \int {\cal D}[\psi^{\dag},\psi] \, {\rm e}^{ -S_{c} -\int_{0}^{\beta} d\tau \int_{0}^{\beta} d\tau' \sum_{\bf K} \psi_{\bf K}^{\dag}(\tau) \Delta_{\bf K}(\tau,\tau') \psi_{\bf K}(\tau') }, 
\label{eq:Z}
\end{equation}
where $S_{c}$ is the action of the cluster, ${\bf K}$ label the cluster momenta, and $\Delta$ is the bath hybridization matrix. 
The self-consistency condition reads 
\begin{equation}
\begin{split}
\Delta(i\omega_{n}) = & \, i\omega_{n} +\mu -\Sigma_{c}(i\omega_{n})  \\
                      & \, -\left[ \sum_{\tilde{k}} \frac{1}{i\omega_{n} +\mu -t_{c}(\tilde{k}) -\Sigma_{c}(i\omega_{n})}\right] ^{-1}.
\end{split}
\label{eq:SCC}
\end{equation}
Here $\Sigma_{c}$ is the cluster self-energy matrix, $t_{c}(\tilde{k})$ is the matrix of hopping in the supercell notation and $\tilde{k}$ runs over the reduced Brillouin zone of the superlattice. 

To solve this problem, we use the numerically exact continuous time quantum Monte Carlo method \cite{Werner:2006,hauleCTQMC}, a finite temperature approach that relies on the Monte Carlo summation of all diagrams obtained from the expansion of the partition function (\ref{eq:Z}) in powers of the hybridization $\Delta$. 
This method does not have errors associated with time discretization or bath parametrization.

%
{\it Phase diagram:} Our results are summarized by the temperature $T$ versus interaction strength $U$ and chemical potential $\mu$ phase diagram shown in Fig.~\ref{fig1}a. 
We performed scans at constant $U$, varying the chemical potential for several temperatures. 
These scans are indicated as cross-sections in the phase diagram and as distinct $T-\mu$ planes in Figs.~\ref{fig1}b to \ref{fig1}e. 
Due to particle-hole symmetry, the phase diagram is symmetric with respect to the $\mu=0$ plane, where the model is half-filled. 
In that plane, we find the results of Ref. \onlinecite{phk} where the $U$ driven first-order metal-insulator transition (MIT) is bounded by the spinodals $U_{c1}(T)$ and $U_{c2}(T)$ (dashed lines) between which two solutions of the cluster DMFT equations, one metallic-like, and one insulating-like, coexist.   
The spinodals end at a second order point $U_{MIT}$. 

We found that the coexistence region extends to {\it finite} values of the chemical potential $\mu$, and is delimited by the spinodal surfaces $\mu_{c1}(U,T)$ and $\mu_{c2}(U,T)$ that end at a critical line (dotted line).  
The intersections between these surfaces and the cross-sections at constant $U$ are displayed in the phase diagram as dark grey (blue) shaded areas. 
For scans at $U=5.6t$ which lie in the region $U_{c1}(T=0) < U < U_{c2}(T=0)$, only the $\mu_{c1}$ line exists. 
For scans at $U=5.8t$ corresponding the region $U_{c2}(T=0) < U < U_{MIT}$, we find coexisting solutions at finite doping between  $\mu_{c1}(T)$ and $\mu_{c2}(T)$. 
As U increases, this region narrows and occurs at lower temperature.

The Mott insulating phase, characterized by a plateau in the occupation at $n=1$, is shown for various constant $U$ planes as light grey (yellow) shaded regions. 
Even though in the region $U_{c1}(T=0) < U < U_{c2}(T=0)$ the Mott insulator can coexist with a metal, as indicated by a different shade of grey (yellow), this is generally not so. 
In all cases displayed, the $\mu_{c2}(U,T)$ surface marks the vanishing of a metallic solution while the $\mu_{c1}(U,T)$ surface does not coincide with the vanishing of the Mott insulating state. 
Doping instead occurs gradually in certain cluster momenta, a phenomenon called `sector-selective transition' \cite{michelEPL, werner8}.
This feature is the coarse grained analogue of a continuous appearance of the Fermi surface with doping, supporting the formation of arcs or pockets in the Fermi surface at small doping. 

More precisely, in the plaquette scheme there are four cluster momenta, $(0,0)$, $(\pi,0)$, $(0,\pi)$ and $(\pi,\pi)$, which can be thought to represent a coarse grained approximation of the Brillouin zone. 
In the region between the Mott insulator and $\mu_{c1}$, the $(\pi,\pi)$ sector remains gapped and its occupation does not change with $\mu$, but the other sectors are gapless (cf filled squares in the $T-\mu$ cuts of Figs.\ref{fig1}b to \ref{fig1}e). 
For $\mu < \mu_{c1}$, all the sectors are gapless. 
The selective metallization of the Mott insulator is not restricted to moderate values of the coupling, but also takes place for large values of $U$.
The transition between the sector selective metal and the metal can be first order or it can occur as a crossover depending on $U,T$ and $\mu$.

\begin{figure}[!ht]
\centering{\includegraphics[width=0.8\linewidth,clip=]{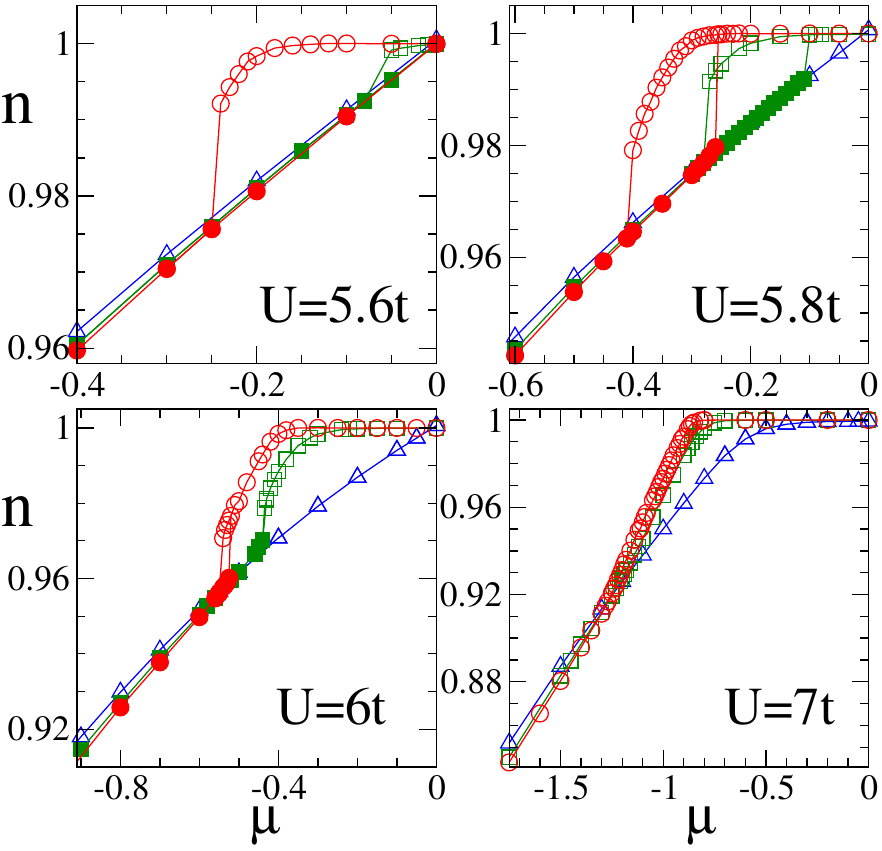}}
\caption{Occupation $n$ versus $\mu$ for different values of interaction strength $U$. The data shown are for temperatures $T/t=1/10, 1/25, 1/50$ (triangles, squares and circles respectively). When two solutions are found to coexist, the solutions obtained following the metallic and the insulating solution are indicated as full and open symbols respectively. The plateau in the occupation at $n=1$ signals the onset of the incompressible Mott state. 
}
\label{fig2}
\end{figure}
{\it First-order transition at finite doping:} The plot of the occupation $n$ as a function of $\mu$, shown in Fig.~\ref{fig2} for different values of $U$ and a wide range of $T$ is the best way to exhibit the first-order transition. 
A plateau in the curves at $n(\mu)=1$ appears above the critical coupling $U_{c1}(T)$ and reveals the onset of the incompressible Mott state.  
The curves $n(\mu)$ display a strong temperature dependence, especially in the phase nearest to the Mott insulator, indicating that the strong correlation effects appear at low energies.

For $U/t=5.6, 5.8, 6$ and low temperatures, we find an hysteresis loop between two solutions, an unambiguous signature of the first order character of the transition.  
The Clausius-Clapeyron equations reveal that the sector selective metal closest to the Mott insulator has smaller entropy and smaller double-occupation than the large doping phase.
The hysteresis loop becomes larger with decreasing T. 
For $U=5.6t$, located in the region $U_{c1}(T=0) < U < U_{c2}(T=0)$, the branch of the hysteresis cycle that corresponds to increasing $\mu$ (i.e. the metallic branch), continues up to half-filling, at $\mu=0$.
In contrast, for $U$ larger than $U_{c2}(T=0)$, namely $U=5.8t$ and $U=6t$, the metallic branch shows an upward jump at finite doping. 
The metallic branch endpoint defines $\mu_{c2}$. 
On the other hand, the upper branch of the hysteresis cycle, obtained by decreasing $\mu$ from the insulator, evolves continuously into the sector selective metal and then undergoes a downward jump at {\it finite doping}. 
This defines $\mu_{c1}$. 

The first order transition occurs between two different metals. Since these phases have the same symmetry, the first-order surface can end at a critical line at finite temperature. 
The $n(\mu)$ curves for $U=7t$ do not show an hysteresis loop but the jump in compressibility $(dn/d\mu)_T$ at the lowest temperature, still visible, suggests that the hysteresis loop is very small and that we are near the critical line. 
In addition, the crossing of the isotherms near $n\sim0.91$ translates into a vanishing expansion coefficient $(dn/dT)_\mu=0$. That feature, in this temperature range, continues to larger doping as $U$ increases (not shown) as a left-over of the critical end line.

{\it Discussion:} A natural way to understand the effects of the short-range spin correlations on the Mott transition, is to compare our cluster DMFT phase diagram, Fig.\ref{fig1}a, that accounts for these magnetic correlations, with that of single-site DMFT, where those correlations are absent. 
In single-site DMFT, the half-filled Mott insulator is connected to a correlated metallic state through a first order transition \cite{rmp}, which can take place as a function of either $U,T$ or $\mu$. The first order surface ends at finite temperature at a critical line \cite{sahana}. 

Short-range spin correlations modify the single-site picture. 
In the $U$-driven MIT, scrutinized in Ref.~\cite{phk}, and visible in our phase diagram along the plane $\mu=0$, the transition is still first order, but the critical value of the Mott endpoint is reduced by roughly a factor of two and the shape of the spinodal lines is modified because the $T=0$ insulating state has no entropy when magnetic correlations are taken into account. 
Our exhaustive calculations in the $T-\mu-U$ space reveal that the short range spin correlations have an additional surprising effect.  
They shift the spinodal surface $\mu_{c1}$ to {\it finite} doping, i.e. the first order transition for $U>U_{c2}(T)$ is between two metals instead of between an insulator and a metal.   

\begin{figure}[!ht]
\centering{\includegraphics[width=0.8\linewidth,clip=]{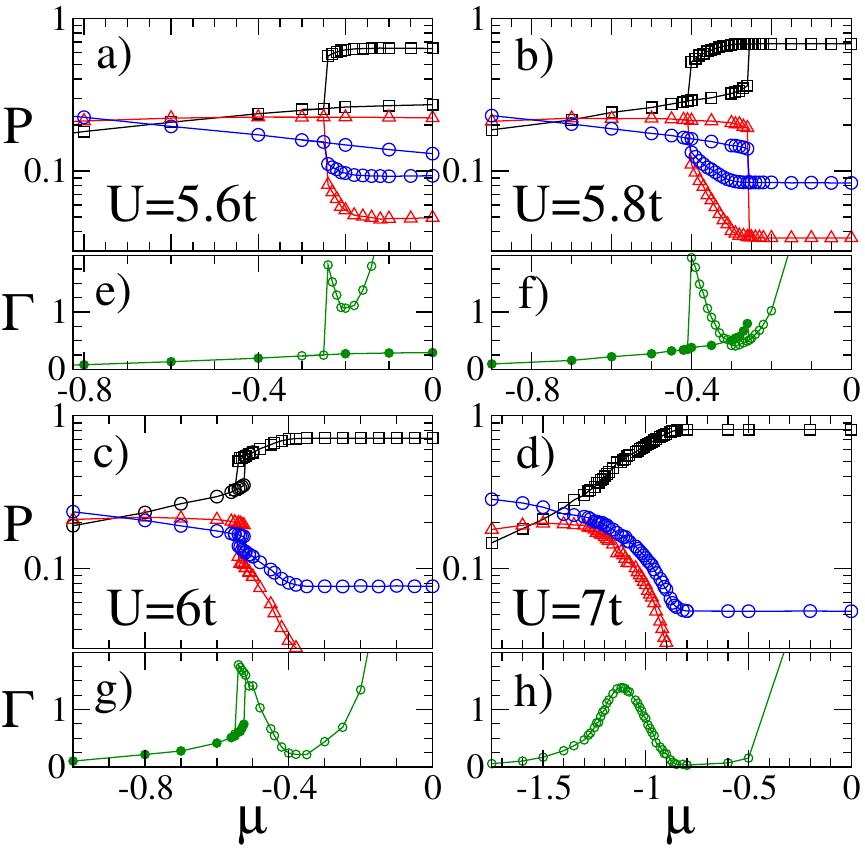}}
\caption{(a) to (d): statistical weight $P$ of the following cluster eigenstates as a function of $\mu$, for several values of $U$, and at the low temperature $T/t=1/50$: 
the singlet $|N=4, S=0, K=(0,0)\rangle$, triplet $|N=4, S=1, K=(\pi,\pi)\rangle$, and doublet $|N=3, S=1/2, K=(\pi,0)\rangle$ (squares, triangles and circles respectively)
where $N$, $S$ and $K$ are the number of electrons, the total spin and the cluster momentum of the cluster eigenstate. 
(e) to (h): scattering rate $\Gamma$=-Im$\Sigma_{(\pi,0)}(\omega\rightarrow 0)$ in the sector $(\pi,0)$ versus $\mu$, at $T/t=1/50$. 
}
\label{fig4}
\end{figure}

Insight on the role of spin correlations can also be obtained from the plaquette eigenstates \cite{hauleDOPING}. We find, at low temperature, that the states with largest probability are the singlet with 4 electrons in the cluster momentum $K=(0,0)$, the half-filled triplet in $K=(\pi,\pi)$, and the doublet with 3 electrons and $K=(\pi,0)$. In Figs.~\ref{fig4}a to \ref{fig4}d we show their statistical weight $P$ as a function of $\mu$. 
At low doping, the half-filled singlet dominates, revealing that the spins are bound into singlets due to the superexchange mechanism. We also find that the hybridization function depends only weakly on the sector, consistent with spin correlations within the plaquette that dominate. 
At large doping (small $\mu$), the charge excitation becomes more important, as expected, so the weight of the doublet at $N=3$ has the highest probability. 

For large values of $U$ (see $U=7t$), at this $T$ the weight of the doublet is larger than the triplet in the whole doping range and no coexistence is found. 
On the contrary, for intermediate values of $U$ (see $U/t=5.6, 5.8, 6$), the doublet and the triplet states compete. 
The coexistence region is associated with this interplay: in the metallic branch of the hysteresis loop, the triplet prevails over the doublet; in contrast, in the insulating branch, it is the doublet that wins over the triplet (similarly to the large $U$ case at low doping). 
Since the doublet at $N=3$ corresponds to charge excitations, and the triplet at $N=4$ to spin excitations, the first order transition is associated to a competition between spin and charge. 
Once the spin-charge competition is removed, at sufficiently large $U$ for our lowest $T$, we are above the critical end line and a smooth crossover takes place. 

Competing charge and spins fluctuations near the Mott transition are a source of scattering, as can be seen from  Figs.~\ref{fig4}e to \ref{fig4}h.  
There we plot the scattering rate $\Gamma(\mu,T/t=1/50)$, estimated from the extrapolation of the imaginary part of the cluster self-energy in the $(\pi,0)$ sector.
In a Fermi liquid, $\Gamma$ as well as the scattering rate at all other cluster momenta, should vanish.
In contrast, leaving aside the insulator, we find a maximum in $\Gamma$ either near the hysteresis loop of the first order transition, or near the critical end line at large $U$ (see for example $U=7t$).
We have verified that as temperature increases, the value of $\Gamma$ at its maximum increases as does its width in doping, reminiscent of the behavior observed around optimal doping in the cuprates.

Several groups, using different cluster methods, already reported an anomalous metallic state at finite doping, linking this to the pseudogap phenomenon \cite{kyung,jarrellNFL, liebsch}, to the selective transition \cite{werner8, michelEPL}, or to a competition between Kondo and superexchange scale \cite{hauleDOPING, hauleCRITICAL,michelEPL}. 
However, those investigations mostly focused on a region where the interaction strength is large. Our contribution is to track the origin of the large scattering rate to the influence of the Mott phenomenon at finite doping, where spin and charge fluctuations compete. 
 
Although the first order transition and its associated maximum in $\Gamma$ occurs at small doping in the $T>0.01t$ range that we can access, its signature in the form of a crossover occurs at progressively larger doping as $U$ increases, reaching the optimal doping point found for $U=12t$ in Ref.~\onlinecite{hauleDOPING}. The crossover, as measured by the maximum scattering rate or by $(dn/dT)_\mu=0$, is caused by the influence of the critical line that moves to progressively lower temperature and higher doping as $U$ increases. Accounting for differences in methods, this crossover caused by critical behavior is also the one detected for $U=8t$ in Refs.~\onlinecite{jarrellNFL,Jarrell:2009}. We cannot strictly exclude that the critical line that ends our first-order surface becomes a quantum critical line or point at some critical $U$ but this possibility is unlikely. 


We conjecture that the quantum critical behavior surmised in high-temperature superconductors \cite{Aeppli,Tallon} is a constant $U$ cut of the low temperature behavior identified in our work as originating from the finite-doping critical end line of the first-order surface between two types of metals. That transition occurs only when $U$ is larger than the critical value $U_{c1}$ for the Mott transition. The influence of Mott physics extends way beyond half-filling.


We are indebted to S. Allen for technical help and to D. S\'en\'echal for useful discussions. This work was partially supported by FQRNT and by the Tier I Canada Research Chair Program (A.-M.S.T.). Computational resources were provided by CFI, MELS, the RQCHP and Compute Canada.


\end{document}